\UseRawInputEncoding
\documentclass[reprint,aps, prb,twocolumn,amsmath,amssymb,nofootinbib,longbibliography]{revtex4-2}
\usepackage{graphicx}
\usepackage{dcolumn}
\usepackage{bm}
\usepackage{color}
\usepackage{lipsum}
\usepackage{hhline}
\usepackage[a4paper]{hyperref}
\usepackage{mwe}
\hypersetup{colorlinks=true,linktoc=all,linkcolor=blue,breaklinks=true,citecolor=blue,urlcolor=blue}
\newcommand{\be}{\begin{equation}}
\newcommand{\ee}{\end{equation}}
\newcommand{\bea}{\begin{eqnarray}}
\newcommand{\eea}{\end{eqnarray}}

\usepackage{natbib}

\begin{document}

\title{Magnon-plasmon coupling  mediated by linear magnetoelectric effect in two-dimensional crystals  with Dzyaloshinskii-Moriya interaction}

\author{Wojciech Rudzi\'nski$^{1}$}
\email{wojciech.rudzinski@amu.edu.pl}
\author{Mirali Jafari$^{1}$}
\author{J\'ozef Barna\'s$^{2}$}
\author{Anna Dyrda\l$^{1}$}

\affiliation{$^{1}$Faculty of Physics, Adam Mickiewicz University in Pozna\'n, ul. Uniwersytetu Pozna\'nskiego 2, 61-614 Pozna\'n, Poland}

\affiliation{$^{2}$Institute of Molecular Physics, Polish Academy of Sciences, ul. Mariana Smoluchowskiego 17, 60-179 Pozna\'{n}, Poland}

\date{\today}
\begin{abstract}
Recently, one can observe a renewed interest in coupling of spin waves (magnons) and collective charge oscillations (plasmons), especially in two-dimensional systems. Several mechanisms of the magnon-plasmon hybridization in ferromagnetic and antiferromagnetic systems have been proposed. Here, we consider another mechanism of magnon-plasmon hybridization, which is based on the linear magnetoelectric interaction. As a specific system we consider a monolayer of vanadium-based diselenide with perpendicular easy-axis magnetic anisotropy and Dzialoshinskii-Moriya interaction. The derived parameter of magnon-plasmon coupling is proportional to the magnetoelectric constant. Assuming for this constant an adequate experimental value, we calculate dispersion relations of the hybridized magnon-plasmon mods. Moreover, we also show that an external electric field normal to the layer (due to a gate voltage) can be used as a tool to tune the magnon modes and this way also hybridized magnon-plasmon coupling. A specific case of magnon-plasmon coupling based on tuning Dzialoshinskii-Moriya interaction is also considered.
\end{abstract}
\maketitle

\section{Introduction}

Elementary excitations in solids (like phonons, plasmons, magnons, and others) and their mutual interactions are of fundamental interests for a proper understanding of many physical properties of materials. In case of bulk 3-dimensional (3D) materials, coupling between various excitations  was a subject  of extensive research for several past decades, especially in case of magnon-phonon and plasmon-phonon interactions. When a  magnetic material is conducting,  it can support both magnetic (magnons) and collective electronic (plasmons) elementary excitations~\cite{Atland,VignaleBook,Gross}.  These two different types of excitations may become coupled if certain conditions are fulfilled. First,  the coupling must be admitted by symmetry of the system. Second, frequencies of the two excitations should be comparable.  In turn, to be observed experimentally, the coupling should be sufficiently strong.  However, there was only a little interest in magnon-plasmon coupling~\cite{Baskaran,JB1,JB2,JB3} until the last decade. This happened because hybridization of magnons and plasmons is difficult to be reached in bulk 3D materials due to a usually large difference in frequencies of these two excitations.
Indeed, this difference in frequencies is especially large in 3D metallic systems, where the bulk plasmon modes appear in the optical frequency range~\cite{VignaleBook}, while the magnon frequencies are usually in the GHz and THz regions in ferromagnets (FM) and antiferromagnets (AFM), respectively~\cite{RezendeJAP2019,akhiezer1968spin}. From this point of view, magnetic semiconductors are more suitable for the observation of magnon-plasmon hybridized states, due to a much lower electron density and thus also lower plasmon frequency. Accordingly, the magnon-plasmon hybridization can be then reached experimentally, as pointed out a few decades ago in Ref.~\cite{JB1}, where also a mechanism of this coupling, based on the spin-orbit interaction associated with localized magnetic moments,  was formulated.
It was also proposed, that the resonant magnon-plasmon coupling can be achieved much easier for surface (interface) magnon and plasmon modes, because such modes have usually  frequencies reduced when compared to the corresponding bulk modes~\cite{JB4}, or in system of magnon and plasmon polaritons. Apart from this, a great attention was paid to
plasmons in semiconducting  materials with strong spin-orbit interaction, where the concept of spin-plasmons was introduced and studied mainly theoretically~\cite{Polini,Kargarian,Magarill,Kushwaha2006,Xu2003,Magarill2001,Wang2005}.

Recently, the interest in coupling between magnons and plasmons  has revived, mainly in the context of coupling between interfacial magnons and plasmons~\cite{PhysRevB.104.094528} or interfacial plasmon and magnon polaritons~\cite{JB4,PhysRevMaterials.6.085201,PhysRevLett.133.156703,PhysRevB.109.214436} in heterostructures. In  Ref.~\cite{PhysRevB.104.094528}, the author has considered coupling of antiferromagnetic magnons to Mooij-Sch\"on plasma waves in a superconducting cavity. The considered coupling appears due to the interface exchange interaction of spin-orbit coupled electrons in a thin superconducting film with localized spins of adjacent antiferromagnet. In turn, in  Refs ~\cite{PhysRevLett.133.156703,PhysRevB.109.214436},  Maxwell equations of classical electrodynamics, together with appropriate boundary conditions, were used  to describe coupled magnon-plasmon polaritons at the interface in  a heterostructure consisting of semiconducting  and ferromagnetic~\cite{PhysRevLett.133.156703}  or antiferromagnetic~\cite{PhysRevB.109.214436} materials. In the former case the authors focused on calculating the reflection spectrum, and the dips found in the spectrum attributed to the mixed magnon-plasmon polaritons. In the latter case, in turn, the description was extended by including the effects due to spin pumping and spin torque described using the  Landau-Lifshitz-Gilbert equation for magnetic dynamics. In Ref.~\cite{PhysRevMaterials.6.085201}
the coupled magnon-plasmon-phonon polaritons were studied in a topological insulator/antiferromagnetic bilayer heterostructure. Within the conventional approach to polariton modes and using the scattering matrix approach, the authors calculated transmission across the structure and the spectra of coupled magnon-plasmon-phonon polaritons.

A new opportunity for the investigation of coupled magnon-plasmon modes appeared with the discovery of  two-dimensional (2D) crystals and  their heterostructures~\cite{Bludov,CostaNL,Katsnelson,Hwang}, where the  plasmon mode is gapless, i.e., its frequency goes to zero for vanishing wavevector~\cite{Platzman,VignaleBook}.  In Refs.~\cite{Bludov}, the authors analyzed coupled magnon-plasmon  polaritons in a heterostructure consisting of a graphene and an insulating antiferromagnet, that are separated by a dielectric layer. Within the classical approach they analyzed coupling between the surface magnon polaritons (supported by the antiferromagnetic layer) and surface plasmon polaritons (supported by grahene), that leads to the mixed magnon-plasmon polaritons. In turn,  it was shown in Ref.~\cite{CostaNL}, that the coupling between magnon and plasmon  polaritons in such a heterostructure can be extremely strong. Another theoretical approach to coupled magnon-plasmon states in hexagonal 2D magnets was proposed in Ref.~\cite{Katsnelson}, where the authors formulated the relevant Hamiltonian including spin, spin-orbit, exchange, and electron-electron interactions. Then, the spectrum of excitations has  been derived from  the  corresponding 'loss' function (inverse of the dielectric function). The dielectric function was also used earlier to evaluate plasmon modes in graphene~\cite{Hwang}.

Recently, an effective mechanism of magnon-plasmon hybridization in 2D materials,  based on  the spin-orbit coupling (SOC) in the subsystem of mobile electrons and the {\it{s}}--{\it{d}}({\it{f}}) exchange interaction between the mobile electrons and the spin subsystem, has been formulated.
This mechanism relies on the spin polarization of conduction electrons due to the electric field associated with plasmon oscillations. Such a non-equilibrium spin density induced by electric field is known to appear when a SOC associated with mobile electrons, e.g. of Rashba and/or Dresselhauss forms,  exists in the system~\cite{Dyakonov,Ivchenko,Edelstein,Wang,Dyrdal,Raimondi}. The plasmon-induced spin density of mobile electrons couples to magnon modes in the spin lattice via the {\it{s}}--{\it{d}}({\it{f}}) exchange interaction. In fact, this mechanism is applicable also to 3D materials.

In this paper we develop a general description of magnon-plasmon coupling in systems which  exhibit linear magnetoelectric properties~\cite{scott,me_data,PhysRevB.109.214435}, and also  host the  Dzyaloshinskii-Moriya interaction (DMI)~\cite{PhysRevLett.35.1017,Moon_2013,Puszkarski_2017}.  The main feature of systems exhibiting linear magnetoelectric effect is the  linear magnetic response to electric field and linear electric response to a magnetic field. This phenomenon is known already for roughly six decades and its static properties have been investigated in many systems~\cite{Freeman,scott,me_data}. Its impact on spin dynamics was analyzed as well~\cite{barnas_JPC1,barnas_JPC2}. The phenomenological parameter $\alpha_{me}$ of linear magnetoelectric coupling  effectively includes contributions from all  relevant microscopic interactions, including also DMI.  There are many experimental data on linear magnetoelectric constant $\alpha_{me}$, and they range from $\alpha_{me}\approx 10^{-4}$ to $\alpha_{me}\approx 10$ in the dimensionless cgs units. Most of these data concern 3D systems. For more information see, e.g., Ref.~\cite{me_data}.
In turn, DMI leads to various interesting phenomena, like noncollinear spin textures (including skyrmions) or nonreciprocal magnon propagation -- for a recent review see~\cite{CAMLEY2023100605}.
The model of magnon-plasmon coupling, proposed in this paper, is based on the magnetic (spin)  polarization, induced  by electric field associated with plasmons due to the linear magnetoelectric interactions.

As a model system we have chosen the vanadium-based dichalcogenides~\cite{chhowalla2013chemistry,zhang2013dimension,ma2012evidence,gao2013ferromagnetism}. First, Vanadium-based dichalcogenides have usually high Curie temperatures, in the range of room temperatures. Second, their magnetic properties, including magnetic anisotropy, exchange and DMI parameters  and others, can be tuned  by an external  strain or by proximity to adjacent layers of other materials~\cite{Fert_Janus_2020}. As a specific material for our analysis we  choose a monolayer of vanadium diselenide, VSe$_2$, with perpendicular magnetic anisotropy. Such a magnetic configuration is important in our case as it allows for tuning the spin wave energy (and thus the magnon-plasmon coupled states) with an external vertical electric field due a gate voltage.
\begin{figure}
\includegraphics[width=\columnwidth]{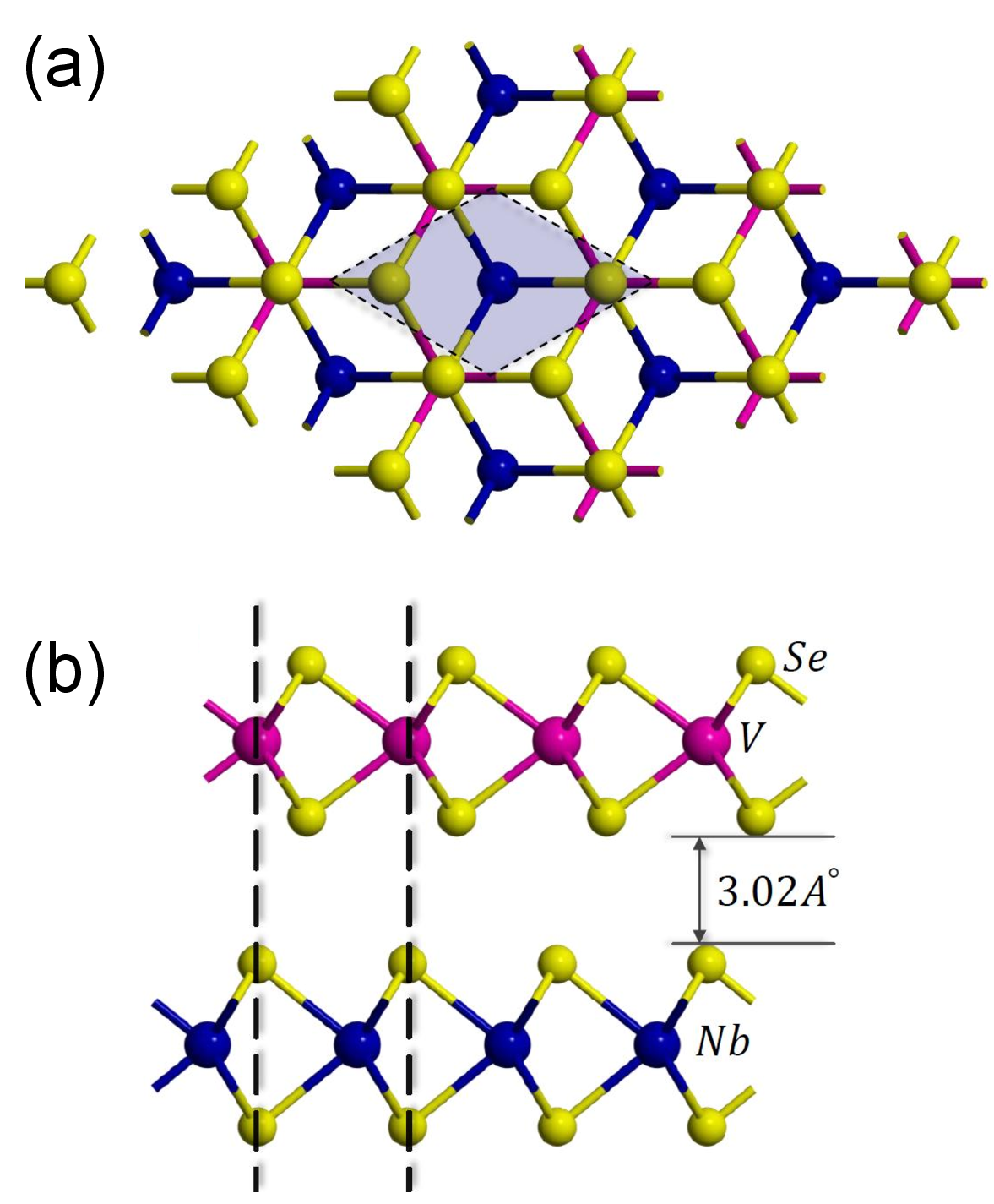}
\caption{Atomic structure of the VSe$_2$ monolayer deposited on a NbSe$_2$ monolayer. Top view (a) and side view (b).}
\label{fig:Fig1}
\end{figure}

As it is known, pristine  monolayers of VSe$_2$ have an in-plane magnetic anisotropy, so the magnetic moments of Vandium atoms in the ground state are oriented in the layer plane. However, when deposited on an another transition-metal dichalcogenide, e.g. on Niobium diselenide, NbSe$_2$ (see Fig.\ref{fig:Fig1}), the magnetic anisotropy can change from the in-plane  to out-of plane (perpendicular) one, as shown already by DFT calculations~\cite{PhysRevB.108.024427}. This possibility, however, is very sensitive to internal parameters (like Coulomb correlations, or spin-orbit coupling) used in DFT calculations. Such a behavior  occurs because NbSe$_2$  is a specific material, with a tendency to form a charge-density-wave (CDW) ground state. Moreover, the transition to perpendicular orientation in a VSe$_2$/NbSe$_2$ is also remarkably sensitive to external strain, as follows from our DFT calculations. For more details on the DFT results and the corresponding parameters
and transport properties, see the Appendix A. We also note, that in general most of the parameters describing 2D materials can be tuned externally, e.g.,   by strain, gate voltage, or proximity effects~\cite{zhang_gate-tunable_2020,kartsev_2020,Jaeshke-Ubiergo_PhysRevB_21,Ebrahimian,jafari2023electronic,stagraczynski2024magnetic,jafari2024effect}.

As we intend to study magnon-plasmon hybridization, the system we need has to display both nonzero magnetization and metallic or semiconducting electronic transport properties. From Appendix A follows that  the VSe$_2$/NbSe$_2$ system under the strain of 2\%  obeys both these conditions.
The band structure shown in Appendix A indicates that the main contribution to electronic states at the Fermi level (and thus also to charge current) comes from the  VSe$_2$ monolayer.
Therefore, our further model considerations will be focused on the conducting VSe$_2$ monolayer with perpendicular magnetic anisotropy. The impact of the NbSe$_2$ monolayer, due to the proximity effect, is included {\it via} the effective exchange parameters (both symmetric and antisymmetric) and by the effective magnetic anisotropy.

\section{Spin waves}

Hamiltonian of the whole magnon-plasmon system includes effectively three terms, $\mathcal{H}= \mathcal{H}_{\rm m} + \mathcal{H}_{\rm pl} +\mathcal{H}_{\rm m-pl}$.
The first term, $\mathcal{H}_{\rm m}$, represents the magnon system, $\mathcal{H}_{\rm pl}$ corresponds to  the system of free plasmons, whereas the last term, $\mathcal{H}_{\rm m-pl}$,  describes coupling between the plasmons and magnons. Let us look now in more details at each component of the Hamiltonian separately, and begin with the magnon Hamiltonian.

Spin waves in VSe$_2$ with easy-plane magnetic anisotropy have been analyzed in our recent work~\cite{RUDZINSKI2023171463}, where the features of spin-wave spectra following from DMI have been considered in detail. However, spin waves in the case of perpendicular easy axis anisotropy have not been investigated yet, so we start our analysis from a brief description of spin waves in this geometry.
The corresponding spin Hamiltonian for the hexagonal spin lattice can be written in a general form as
\begin{equation}
    \mathcal{H}_{\rm m}=\mathcal{H}_{ex}+\mathcal{H}_{A}+\mathcal{H}_{DM},
\end{equation}
where the first term describes the  exchange coupling between localized spins,
\begin{equation}
    \mathcal{H}_{ex}=J_1\sum_{\mathbf{r},{\bm \delta}}\mathbf{S}_{\mathbf{r}}\cdot\mathbf{S}_{\mathbf{r}+\bm{\delta}} + J_2\sum_{\mathbf{r},{\bm \delta}^\prime}\mathbf{S}_{\mathbf{r}}\cdot\mathbf{S}_{\mathbf{r}+\bm {\delta}^\prime}.
\end{equation}
The summation over $\mathbf{r}$ denotes here the summation over all lattice sites, while that over  $\bm \delta$ ($\bm \delta^\prime$) denotes the summation  over the nearest  (next-nearest) neighbors, with the vectors $\bm \delta$ ($\bm \delta^\prime$) joining a  given lattice site to its  nearest (next-nearest) neighbors. The parameters $J_1$ and $J_2$ are the corresponding exchange integrals (positive for ferromagnetic coupling and negative for antiferromagnetic one).  The second term in Eq.(2) stands for the magnetic anisotropy, and has the standard form~\cite{PhysRevB.108.024427},
\begin{equation}
    \mathcal{H}_{A}=-\frac{D_z}{2}\sum_{\mathbf{r}}\Big(S_{\mathbf{r}}^z\Big)^2,
\end{equation}
where the the corresponding anisotropy parameter is positive, $D_z>0$, the for out-of plane easy axis.  The third term in Eq.~(2)  represents the DMI~\cite{Fert_Janus_2020},
\begin{eqnarray}
   \mathcal{H}_{\rm DM}=-\sum_{\mathbf{r},\mathbf{ \bm \delta}}\mathbf{D}_{\mathbf{r},\mathbf{r}+{\bm {\delta}}}\cdot(\mathbf{S}_{\mathbf{r}}\times\mathbf{S}_{\mathbf{r}+{\bm \delta}})\nonumber\\
   -\sum_{\mathbf{r},\mathbf{ \bm \delta^\prime}}\mathbf{D}^\prime_{\mathbf{r},\mathbf{r}+{\bm {\delta}^\prime}}\cdot(\mathbf{S}_{\mathbf{r}}\times\mathbf{S}_{\mathbf{r}+{\bm \delta}^\prime}).
\end{eqnarray}
The Dzyaloshinskii-Moriya vectors $\mathbf{D}_{\mathbf{r},\mathbf{r}+{\bm \delta}}$ can be written in the form~\cite{Fert_Janus_2020},
\begin{equation}
   \mathbf{D}_{\mathbf{r},{\mathbf{ r }}+ {\bm \delta}}=d_{\parallel}({\mathbf{\hat{u}}_{\mathbf{r},\mathbf{r}+\bm{\delta}}\times}\mathbf{\hat{z}})+d_{\perp}\xi _{\mathbf{r},\mathbf{r}+\bm{\delta}}\mathbf{\hat{z}},
\end{equation}
where $\mathbf{\hat{u}}_{\mathbf{r},\mathbf{r}+\bm{\delta}}$ is the unite vector from site $\mathbf{r}$ to site $\mathbf{r}+\bm{\delta}$, $\mathbf{\hat{z}}$ is a unit vector along the axis $z$ (normal to the layer), and $\xi _{\mathbf{r},\mathbf{r}+\bm{\delta}}= -\xi _{\mathbf{r}+\bm{\delta},\mathbf{r}}=\pm 1$, whereas $d_{\parallel}$ and $d_{\perp}$ are constants.  The Dzyaloshinskii-Moriya vectors for DMI between next-nearest neighbors can be written in a similar form, with $d^\prime_{\parallel}$ and $d^\prime_{\perp}$  as the relevant constants.
As the pristine monolayer includes the inversion symmetry center, the DMI disappears upon summation over all lattice sites. However, it may be induced externally by proximity effect due to an adjacent layer (like in the case of VSe$_2$ on NbSe$_2$), external strain, or gate voltage.

Generally, all the parameters in the above model Hamiltonian  include the effects due to strain, proximity, and gating. The effects due to strain and proximity are dominant and will be considered as  fixed in the following description. In turn,  the gate voltage can be used to  tune moderately  these parameters, and we will include the corresponding contributions to the DMI parameters and anisotropy constant more explicitly, Thus, in case of tuning by vertical electric field $E_z$, the vertical DMI parameter $d_\perp$ should  be replaced by an effective one,  $\tilde{d}_\perp$, which can be written as
 $\tilde{d}_\perp =d_\perp +(\delta d_\perp/\delta E_z)\, E_z $. Here, the second term describes the contribution to the DMI parameter due to the vertical electric field. In the following, we introduce the notation $(\delta d_\perp/\delta E_z) E_z \equiv E_z^\perp$, so  one can write $\tilde{d}_\perp =d_\perp + E_z^\perp$. In the same way one can  write $\tilde{d}_\perp^{\prime} =d_\perp^{\prime} + E_z^{\perp '}$, associated with the  next-nearest-neighbors, and $\tilde{D}_z=D_z + E_z^D$ for the anisotropy parameter.  Note,  $E_z^\perp$, $E_z^{\perp '}$ and $E_z^D$ are  expressed in energy units, and they are different in general. Owing to these terms, spin wave energy can be tuned externally by a gate  voltage.

The following theoretical considerations are limited to a collinear magnetic ground state, which exists when the magnetic anisotropy is sufficiently strong to overcome the canting tendency due to DMI, and  therefore stabilizes the collinear configuration. This assumption remarkably simplifies the considerations. In a general case, however, one needs to consider a noncollinear ground state~\cite{PhysRevLett.35.1017,Puszkarski_2017}. In the system considered here, the anisotropy is relatively large and obeys the above  requirement. Accordingly, we use the procedure described in more details in Ref.~\cite{RUDZINSKI2023171463}.
Upon the Hollstein-Primakoff and Fourier transformations, the magnon Hamiltonian $\mathcal{H}_{\rm m}$ in the linear spin-wave approximation can be written in the form,
\begin{equation}
            \mathcal{H}_{\rm m}=\sum_{{\bf k} }\varepsilon_{m} ({\bf k})\,b^{\dagger}_{\bf k}b_{\bf k} \, ,
\end{equation}
where $\varepsilon_{m} ({\bf k})$ is the spin-wave energy,
\begin{eqnarray}
\varepsilon_{m} ({\bf k}) = 2S [(D_z+E_z^D)+J_2(\gamma^\prime_{{\bf k}}-6) - J_1(\gamma_{{\bf k}}-6) \nonumber \\
+ p(d_\perp +E_z^\perp )\gamma_{\rm DMI}+p(d_\perp^{'}  +E_z^{\perp '})\gamma^\prime_{\rm DMI} ].
\end{eqnarray}
Here, $p=\pm 1$ distinguish between two opposite ground-state spin orientations, while the nearest- and next-nearest-neighbor  structural factors are
\begin{eqnarray}
\gamma_{{\bf k}}=2[\cos (k_xa) +2\cos (k_xa/2) \cos (\sqrt{3}k_ya/2)],\nonumber\\
\gamma^\prime_{{\bf k}}=2[\cos (\sqrt{3} k_ya) +2\cos (3k_xa/2) \cos (\sqrt{3}k_ya/2)],
\end{eqnarray}
for the exchange coupling, and
\begin{eqnarray}
\gamma_{{\rm DMI},{\bf k}}=\sin (k_xa) +2\sin (k_xa/2) \cos (\sqrt{3}k_ya/2),\nonumber\\
\gamma^\prime_{{\rm DMI},{\bf k}}=\sin (\sqrt{3} k_ya) +2\sin (3k_xa/2) \cos (\sqrt{3}k_ya/2),
\end{eqnarray}
for the DMI coupling.
Interestingly, the spin-wave energy  depends on the vertical components of DMI ($d_\perp$ and $d^\prime_\perp$), while it is independent of the planar DMI components ($d_\parallel$ and $d^\prime_\parallel$).

\section{Plasmons and magnon-plasmon coupling}

\subsection{Plasmons}

Excitations in the corresponding electronic subsystem include single-electron excitations and collective excitations, i.e. plasmons. Here, we are interested in the latter ones. By introducing collective coordinates for the long-range part of Coulomb interactions, it has been  shown long time ago \cite{Pines, Gross}, that the Hamiltonian of electrons with Coulomb correlation can be transformed into a Hamiltonian that includes three terms. One term describes short-range interacting electron liquid, the second describes free plasmons, and the third term describes electron-plasmon interaction \cite{Pines, Gross}. The latter term leads to Landau damping of plasmons by creating electron-hole pairs~\cite{Hwang}. In the case under consideration,
the mixed magnon-plasmon states occur in narrow  wavevector ranges, and we assume that the Landau damping does not appear  in these regions. Thus, we disregard here the Landau damping and  assume undamped plasmons,
\begin{equation}
\mathcal{H}_{\rm pl}=\sum_{{\bf k} }\hbar \omega_{\rm pl}({\bf k}) a^{\dagger}_{\bf k}a_{\bf k} \equiv \sum_{{\bf k} } \varepsilon_{\rm pl}({\bf k}) a^{\dagger}_{\bf k}a_{\bf k} ,
\end{equation}
where $a^{\dagger}_{\bf k}$ ($a_{\bf k}$) is the creation (annihilation) operator of a plasmon with the  wavevector ${\bf k}$ and frequency $\omega_{\rm pl}({\bf k})=\omega_{\rm pl}(k)$, while $\hbar \omega_{\rm pl}(k)= \varepsilon_{\rm pl}({k})$ is the plasmon energy. Note, the plasmon energy is isotropic, i.e.  it depends only on the absolute value $k$ of the wavevector. However, we emphasize that in the energy regions, where Landau damping occurs, the picture of independent plasmons breakes down due to their strong damping.

The plasmon dispersion relation  in 2D systems has the general form~\cite{VignaleBook,Polini0,Polini,Maslov}:
\begin{equation}
\omega_{\rm pl}(k) \simeq \sqrt{\frac{2\pi n e^2}{m}k} \;.
\end{equation}
Here, $n$ is the areal electron density, $e$ is the electron charge, while $m$ is the  effective electron mass. In a general case, the latter can be larger or smaller than the electron rest mass, $m_0$, and in 2D materials it is mostly smaller than $m_0$.
Importantly, the plasmon dispersion in 2D systems is gapless, $\omega_{\rm pl}(k \rightarrow 0)=0$, contrary to the 3D case, where an intrinsic large gap exists in the plasmon spectrum.

The plasmon excitations are associated with electric field. In two-dimensional system and in the quantum limit, the electric field is given by the formula~\cite{Deigen},
\begin{align}\label{electric}
{\bm E} =\frac{2\pi ne}{\epsilon} \left(\frac{\hbar}{2 A n m}\right)^{1/2} \sum_{\bf k}\frac{{\bf k}}{ \omega^{1/2}_{\rm pl}} (a^{\dagger}_{-\bf k}-a_{\bf k})e^{i{\bf {k}}\cdot {\bf {r}}},
\end{align}
where $A$ is the system area and $\epsilon$ is the material dielectric constant.
Note, the electric field is expressed in terms of plasmon annihilation and creation operators.

\subsection{Magnon-plasmon  coupling {\it via} linear magnetoelectric interaction}

When a system has low symmetry, electric field can induce magnetic polarization, while magnetic field  can polarize electrically the system. In the linear limit, the induced magnetic (electric) polarization is linear in driving electric (magnetic) field. This phenomenon, referred to as the linear magnetoelectric effect, is known since about six decades. The corresponding magnetoelectric energy per unite square (in 2D systems) can be then written as $E_{me} =\alpha_{me} {\bf M}\cdot {\bf E}$, where $\bf M$ is the magnetization per unite square, $\bf E$ is the electric field, while $\alpha_{me} $ is the corresponding parameter of magnetoelectric coupling. In a  general form, this energy can take a more complex form with nondiagonal terms. Here, however, we assume it in a simple isotropic and diagonal form.

To write this energy in terms of local  spin operators, we take into account  the relations, ${\bf M}= (1/A)\sum_i  g\mu_B {\bf S}_i$, where $g$ and $\mu_B$ denote the Lande factor and Bohr magneton, respectively, while ${\bf S}_i$ is the operator of a spin localized at the $i$-th lattice node.
The corresponding Hamiltonian of the linear magnetoelectric coupling of localized spins and electric field due to plasmons can be then written as
\begin{equation}
\mathcal{H}_{me} =\alpha_{me} g\mu_B   \sum_i E_{x,i} S_{x,i},
\end{equation}
where $E_{x,i}$ is the electric field of plasmons [see Eq.(12)] at site $i$, and without loss of generality we assumed here plasmons propagating along the axis $x$. The summation over $i$ in the above equation is over all the lattice sites in the system.

When expressed in terms of magnon and plasmon annihilation and creation operators, the above Hamiltonian takes the following form
\begin{equation}
\mathcal{H}_{\rm m-pl}  =\sum_{\bf k} C_{\bf k}  (a^+_{-\bf k}b^+_{\bf k} + a_{-\bf k}b_{\bf k}
-a^+_{\bf k} b_{\bf k} - a_{\bf k} b^+_{\bf k}),
\end{equation}
where $C_{\bf k}$ is the parameter of magnon-plasmon coupling,
\begin{equation}
C_{\bf k} =  \alpha_{me} g\mu_B \sqrt{An_s}\beta \sqrt{\frac{S}{2\omega_{pl}}}k_x,
\end{equation}
with $n_s$ denoting the areal concentration of localized spins and $\beta = (2\pi ne/\epsilon)\sqrt{\hbar /2Anm}$. Note, the Hamiltonian (14) includes both, resonant and non-resonant terms. The former terms are responsible for hybridization, while the latter ones for virtual creation (annihilation) of magnon-plasmon pairs.

The parameters $\alpha_{me}$ can be taken from experimental data. These parameters are known for many materials, however mainly for 3D limit. Magnitudes of these parameters cover a relatively broad range, depending on the type of materials. For more information on the experimental values of $\alpha_{me}$ see for instance Refs~\cite{me_data,scott}. To calculate the magnon-plasmon coupling parameter from Eq.(15), we need to convert these parameters
to the cgs system of units, where the parameters $\alpha_{me}$ are dimensionless. Then, Eq.(15) gives
$C_{\bf k}$ is in the cgs units of energy. For convenience, in the following $C_{\bf k}$ will be expressed in meV.

To find the dispersion relations of the coupled magnon-plasmon modes, we need to diagonalize the bosonic  Hamiltonian (14). To do this we use the  procedure based on the Bogoliubov transformation, as described, e.g., in Refs~\cite{RezendeJAP2019,White}. At the end, we find the following dispersion relations for the  magnon-plasmon hybrid modes,
\begin{eqnarray}
\label{FM-plasmon}
\varepsilon^{1,2}_{\rm{m-pl}}({\bf k}) =\frac{1}{\sqrt{2}}\left\{\varepsilon_{\rm pl}^2({\bf k})+\varepsilon_{\rm m}^2({\bf k}) \nonumber \right. \\
\left.\pm \sqrt{(\varepsilon_{\rm pl}^2({\bf k})-\varepsilon_{\rm m}^2({\bf k}))^2+16 |\mathcal{C}_{\bf k}|^2 \varepsilon_{\rm pl}({\bf k}) \varepsilon_{\rm m}({\bf k})}\right\}^{1/2}.
\end{eqnarray}
Obviously, in the absence of magnon-plasmon coupling, $C_{\rm k}=0$, the above relations reduce  to those for decoupled magnon and plasmon modes.

\subsection{Magnon-plasmon  coupling {\it via} DMI}

The phenomenological magnetoelectric parameter $\alpha_{me}$ includes effectively contributions from all relevant microscopic interactions.
A specific contribution originates from
DMI and its sensitivity to electric field. More specifically, the mechanism is based on the modulation of $d_\parallel$
by the electric field of plasmons. As a result, we obtain the effective Hamiltonian of magnon-plasmon interaction in the form of Eq.(14), where the relevant coupling parameter takes now the form,
$C_{\bf k}=  -(\zeta_{\bf k} +\zeta^\prime_{\bf k})$, with
\begin{eqnarray}
\zeta_{\bf k} = d_\parallel
\alpha\beta S \sqrt{\frac{S}{\omega_p}}\left\{2\sqrt{3}k_x\cos(k_xa/2)\sin (\sqrt{3}k_ya/2)\right. \nonumber \\
+\left.i\sqrt{2}k_y[\sin (k_xa/2) \cos (\sqrt{3}k_ya/2)+\sin (k_xa)]\right\} \nonumber \\
\end{eqnarray}
and
\begin{eqnarray}
\zeta^\prime_{\bf k} = d^\prime_\parallel
\alpha^\prime\beta S \sqrt{\frac{S}{\omega_p}}\left\{i3\sqrt{2}k_y\sin (3k_xa/2) \cos (\sqrt{3}k_ya/2)\right. \nonumber \\
+\left.\sqrt{6}k_x\,[\cos(3k_xa/2)\sin (\sqrt{3}k_ya/2) +\sin (\sqrt{3}k_ya)]\right\}. \nonumber \\
\end{eqnarray}
Here,  $\alpha $ and $\alpha^\prime $  describe electric-field-induced modification of the in-plane DM parameter $d_\parallel$, and are defined by the relations
$\alpha =
(\delta d_\parallel /\delta E )$ and $\alpha^\prime =(\delta d^\prime_\parallel /\delta E )$. In the limit od small wavevectors, these coupling parameters are linearly dependent on $k$, similarly as the coupling parameters derived based on the phenomenological form of the magnetoelectric interaction. Some difference between the two forms of magnon-plasmon coupling parameters appears  due to structural factors at longer wavevectors.
However, we are not aware of relevant experimental data on the parameters $\alpha$ and $\alpha^\prime$.
Therefore,  the following numerical results will be shown for the magnon-plasmon coupling based on the linear magnetoelectric parameters. This is reasonable as the magnetoelectric parameters are more general and  include contributions from all relevant microscopic interactions. Moreover, the derived magnon-plasmon Hamiltonians have the same forms and differ only in the coupling parameter. Consequently, numerical results on magnon-plasmon coupled modes obtain in the two models are qualitatively similar, and differ  only by the magnitudes of  the coupling parameter, with the results obtained from linear magnetoelectric interaction more reliable, as being more general.

\section{Numerical results}

\begin{figure*}[t]
\includegraphics[width=0.97\textwidth]{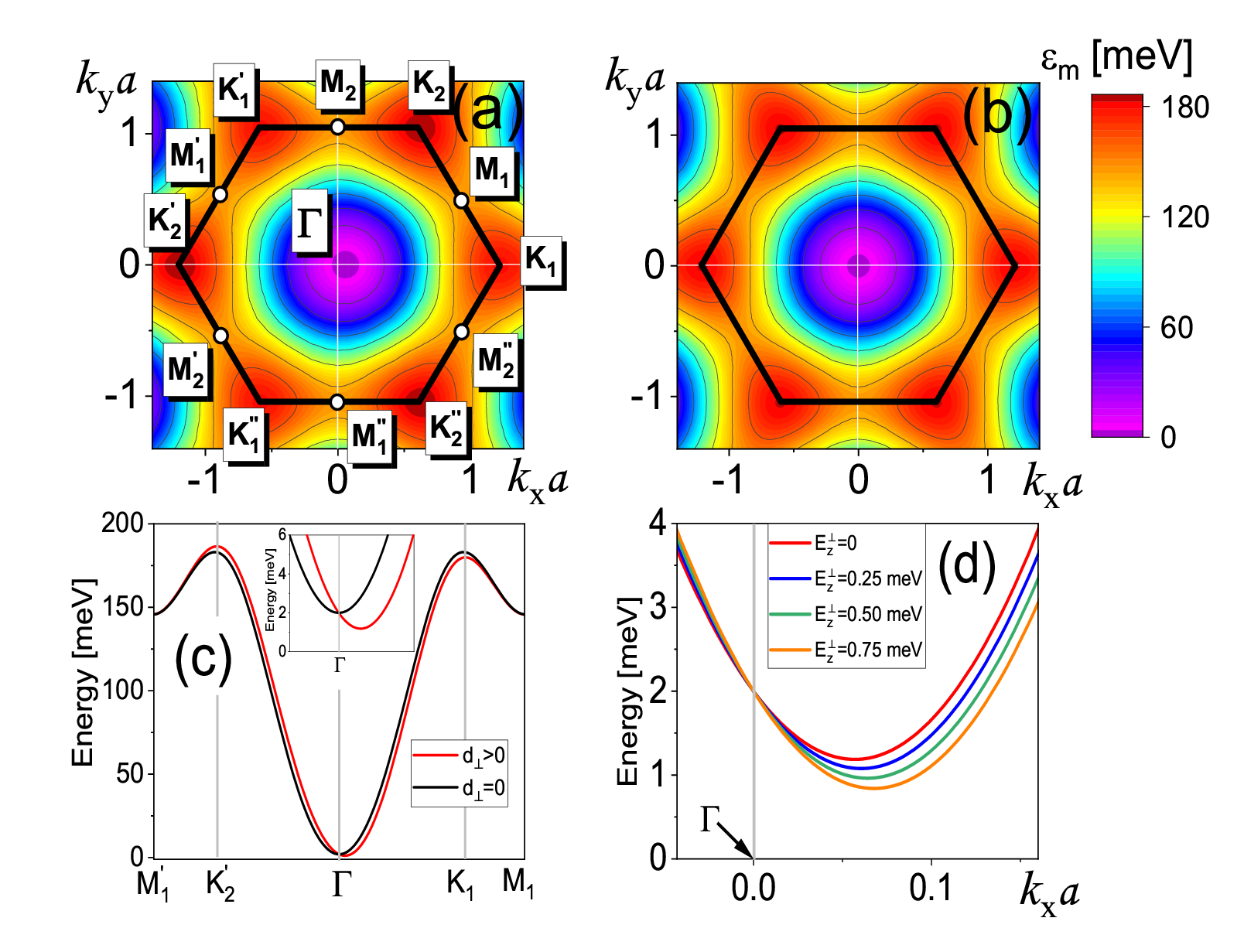}
\caption{Energy of spin waves in the monolayer of VSe$_2$ with perpendicular ground-state spin configuration. (a,b) Energy maps of spin waves in the whole Brillouin zone in the  presence of DMI (a), and  in the absence of DMI ($d_\perp = d^\prime_\perp =0$) (b). (c) Dispersion curves of spin waves in the presence of DMI  (red curves) and in the absence of DMI (black curves). The inset shows the low-energy part of the spectrum. In (a-c) the results are for the absence of external gating. (d) The low-energy part of the spin wave spectrum for the gate fields, $E_z^\perp$, as indicated, while $E_z^D=0$. For all figures, the other parameters are as described in the main text.}
\label{fig:Fig2}
\end{figure*}

For numerical calculations we assume that the localized spins  of VSe$_2$ correspond to the  spin number $S=1/2$, and also
assume a well defined spin polarization in the ground state, corresponding to $p=-1$.
Other material parameters will be taken from DFT calculations for a strained (2\%) monolayer of VSe$_2$ deposited on a monolayer of NbSe$_2$ (see Fig.\ref{fig:Fig1} and also Appendix A). The strain and  proximity to NbSe$_2$ assure perpendicular magnetic anisotropy. When adapted to the assumed  model, the parameters obtained from DFT simulations are: $a=3.47 \AA$  (corresponding to $n_s=1.9 \times 10^{15}$ cm$^{-2}$), $J_1=20.12$ meV, $J_2=2.16$ meV,  $D_z=2$ meV, $d_\perp =3.82$ meV, $d^\prime_\perp =0.186$ meV.
The parameters $d_\parallel$ and $d^\prime_\parallel$ do not enter the formula for spin wave energy in the collinear ground state, and therefore are not used  in the numerical calculations of spin waves and magnon-plasmon coupling {\it via} the linear magnetoelectric coupling.  They would be required for calculations of the magnon-plasmon coupling {\it via} DMI, which however are not performed here.
To calculate plasmon modes and magnon-plasmon coupling parameter, we assume the dielectric constant $\epsilon =1$ and  the areal  concentration of electrons $n= 10^{12} {\rm cm}^{-2}$.  If not stated otherwise, all these parameters will be used in the following numerical calculations. Other parameters used in numerical calculations will be provided where necessary.

\subsection{Spin waves}

According to Eqs (7-9), the spin wave energy depends on $d_\perp$ and $d^\prime_\perp$ (modified by external gate field, in general).   We remind, that the derived formulas are  valid in the linear spin wave approximation and in the limit of collinear ground state, assumed in the description. Let us begin with the presentation of numerical results on pure spin waves, i.e., spin waves decoupled from plasmons. Therefore, we exclude here the magnon-plasmon coupling, assuming $\alpha_{me}=0$.
Before moving to details on the spin wave modes, we make first a general comment on the notation. The Brillouin zone of the pristine VSe$_2$ is hexagonal, with two nonequivalent  Dirac points K$_1$ and  K$_2$. However, to tune the system parameters, e.g. the magnetic anisotropy and DMI constants, one needs to apply externally either strain or electric field (gate voltage). The structural symmetry becomes then changed, and the exact Brillouin zone also differs from the hexagonal ones. However, we keep the initial hexagonal Brillouin zone, but distinguish the three Dirac points K$_1$ and mark them as K$_1$, K$^\prime_1$, K$^{\prime\prime}_1$.  Similar modification also holds for the Dirac points K$_2$. This is necessary to distinguish spin waves for different paths in the Brillouin zone. This difference appears due to  DMI, and vanishes when DMI becomes zero.  Adequately, we also distinguish different points M in the Brillouin zone, as indicated in Fig.\ref{fig:Fig2}(a).

Figure~\ref{fig:Fig2}(a) presents spin waves in the presence of DMI, while Fig.~\ref{fig:Fig2}(b) in the absence of DMI. Both Fig.~\ref{fig:Fig2}(a) and Fig.~\ref{fig:Fig2}(b), show two-dimensional maps of spin wave energy in the whole hexagonal Brillouin zone. As a result of DMI, the spin-wave energy minimum is shifted away from the point $\Gamma$ (towards the point K$_1$). This shift is clearly visible in Fig.~\ref{fig:Fig2}(a), and also in  Fig.~\ref{fig:Fig2}(c), which  shows the dispersion curves along the path M$_1^\prime \to$ K$_2^\prime  \to \Gamma \to$ K$_1 \to$ M$_1$.
The shift of the spin wave energy minimum due to DMI is also clearly visible in the inset to Fig.~\ref{fig:Fig2}(c), which presents the spectrum around the point $\Gamma$. No such a shift appears in the limit of vanishing DMI, see Fig.~\ref{fig:Fig2}(c) (black lines) and also Fig.~\ref{fig:Fig2}(b).

As mentioned above, the energy of spin waves can be modulated by tuning the DMI parameters, $d_\perp$ and $d^\prime_\perp$, with an external electric field (gate voltage), see Eq.(7). To show this we include the dominant contribution originating from the nearest neighbors, $E_z^\perp$, while omit the contribution due to next-nearest-neighbors, $E_z^{\perp '}=0$. In Fig.\ref{fig:Fig2}(d) we show the low-energy spin-wave spectra for indicated  values of $E_z^\perp$, and compare them with the spectrum in the absence of  gating field, $E_z^\perp =0$.
This figure clearly shows that DMI shifts the minimum of spin-wave energy away from the $\Gamma$ point and also to lower energies (for the assumed parameters).
Here we note, that gate field also may tune the magnetic anisotropy parameter $D_z$, and effectively shift the spin-wave spectrum  up/down in energy. However, in Fig.\ref{fig:Fig2}(d) we assumed $E_z^D=0$ and show modulation due to nonzero $E_z^\perp$ only.

\subsection{Magnon-plasmon hybridization}

\begin{figure}
\includegraphics[width=0.96\columnwidth]{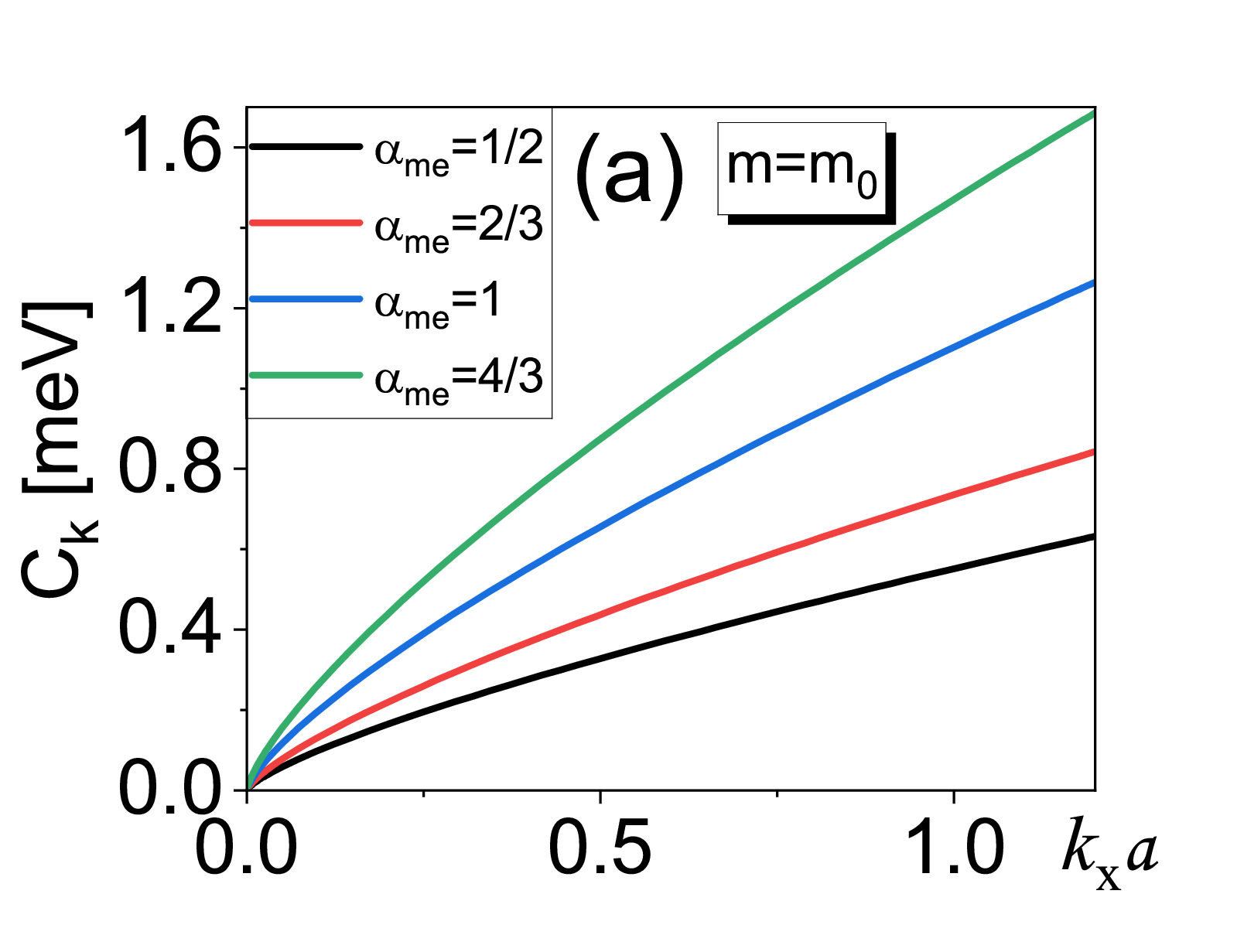}
\includegraphics[width=0.96\columnwidth]{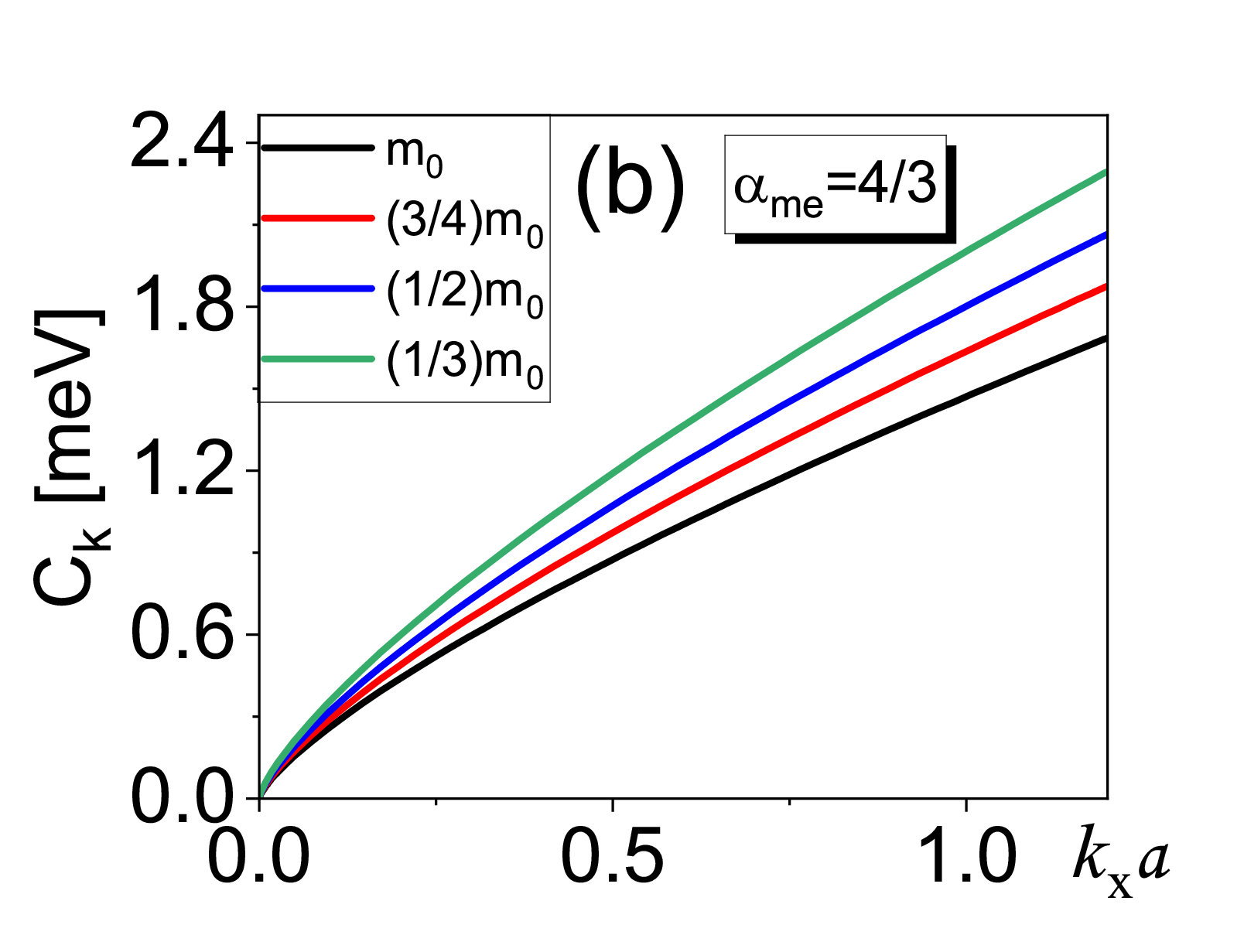}
\caption{Variation of the magnon-plasmon coupling parameter with $k_xa$ in the range from the  Brillouin zone center (point $\Gamma$) to the Brillouin zone boundary (point K$_1$) along the $k_x$ direction. In the part (a)  different curves correspond to the indicated values of parameter $\alpha_{me}$ and for the effective electron mass $m = m_0$. In turn, different curves in (b)  correspond to indicated values  of the effective mass $m$ and constant value of $\alpha_{me} =4/3$. All other parameters assumed in  (a) and (b) are described in the text.}
\label{fig:Fig3}
\end{figure}

Frequency of the plasmon excitations in 2D systems is  described by Eq.~(11). According to this relation, the corresponding plasmon dispersion curves are simple and behave with the wave vector $k$ as $\sqrt{k}$, i.e.,  the dispersion curves are gapless, with zero energy at $k=0$. Assume now the magnon-plasmon coupling is admitted, $\alpha_{me} \ne 0$. As we do not know experimental (or numerical) values of this parameter for the system under consideration,  for numerical calculations of $c_k$ we assume $\alpha_{me}=1/2, 2/3, 1, 4/3$, i.e., values that are somewhere in between the maximal and minimal values of the experimentally measured data~\cite{me_data}. Specifically, for calculating the hybridized magnon-plasmon states we assume  $\alpha_{me} = 1$.

The magnon-plasmon coupling Hamiltonian is given by Eq.(14), and consists of a resonant term responsible for hybridization and a nonresonant term. In the system considered here, there are two modes, which become hybridized around the crossing points of the corresponding dispersion curves of noninteracting modes. This interaction appears as an anticrossing behaviour of the dispersion curves (also referred to as mode repulsion).
Positions of the crossing points depend  on the shape and specific features of the dispersion relations. First of all, the plasmon energy in 2D systems vanishes in the Brillouin zone center, and then grows with the wavevector $k$  as $\sqrt{k}$. In turn, due to magnetic anisotropy, the magnon energy is finite (though usually small) at the $\Gamma$ point. Following this, the magnon and plasmon dispersion curves usually cross each other in the small energy (small wavevector) regime. In turn, according to Refs.~\cite{PhysRevLett.133.156703,LUO2013351,LUO2013351},  the plasmon energy in many 2D system is about few tens (up to a few hundreds) of meV  at the Brillouin zone boundary.
Consequently, in many 2D systems the magnon and plasmon energies are comparable far from the Brillouin zone centre, so a crossing point  of the magnon and plasmon dispersion curves also may appear in the high energy regime, in general.
Such a situation takes place
in  Vanadium-based dichalcogenides, considered here.
As we have already mentioned above, there are some important features of two-dimensional materials, that may facilitate  tuning  the resonant magnon-plasmon  interaction. First, one can  tune the spin wave energy by an external gate voltage, see Eq.(8), that can modulate the magnetic anisotropy constant as well as the DMI parameters. Second, as the plasmon energy varies with the effective electron mass $m$ as $\sqrt{1/m}$, one can use this dependence to modulate the plasmon energy by tuning the electronic band structure and thus the effective electron mass. This gives an additional possibility to reach and/or tune the resonant magnon-plasmon coupling.

The key parameter describing the mode repulsion is the magnon-plasmon coupling parameter $c_k$, see Eq.~(15). This parameter vanishes in the point $\Gamma$, and then grows with the distance from the Brillouin zone center. In Fig.\ref{fig:Fig3} we show its variation with $ka$ for several different situations. In Fig.\ref{fig:Fig3}(a) we show the coupling parameter $c_k$ for a fixed electron effective mass $m$ equal to the rest electron mass $m_0$, and for the four indicated values of the magnetoelectric parameter $\alpha_{me}$. In turn, in Fig.\ref{fig:Fig3}(b), the coupling parameter is shown for a fixed magnetoelectric parameter $\alpha_{me}$ and for four values of the effective electron mass, as indicated. The coupling parameter $c_k$ reaches a few meV at the zone boundary, while it is much smaller near the Brillouin zone center.
%Note, this coupling parameter $c_k$ is probably overestimated due to rather large values of the magnetoelectric parameter $\alpha_{me}$.
It is worth to note, that measuring the dispersion relations of the  coupled magnon-plasmon states gives the possibility to determine the coupling parameter $c_k$. 
\begin{figure}
\includegraphics[width=0.96\columnwidth]{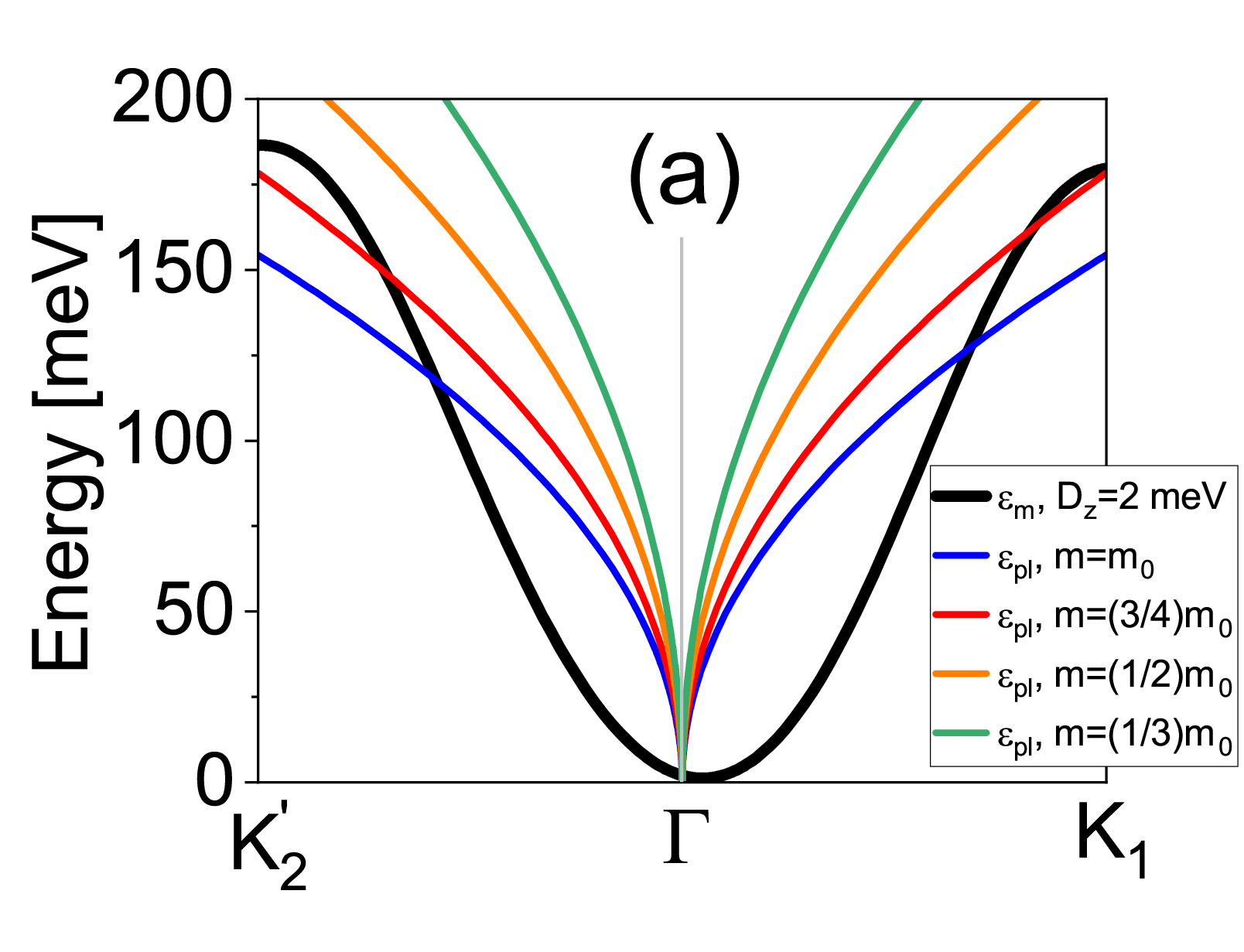}
\includegraphics[width=0.96\columnwidth]{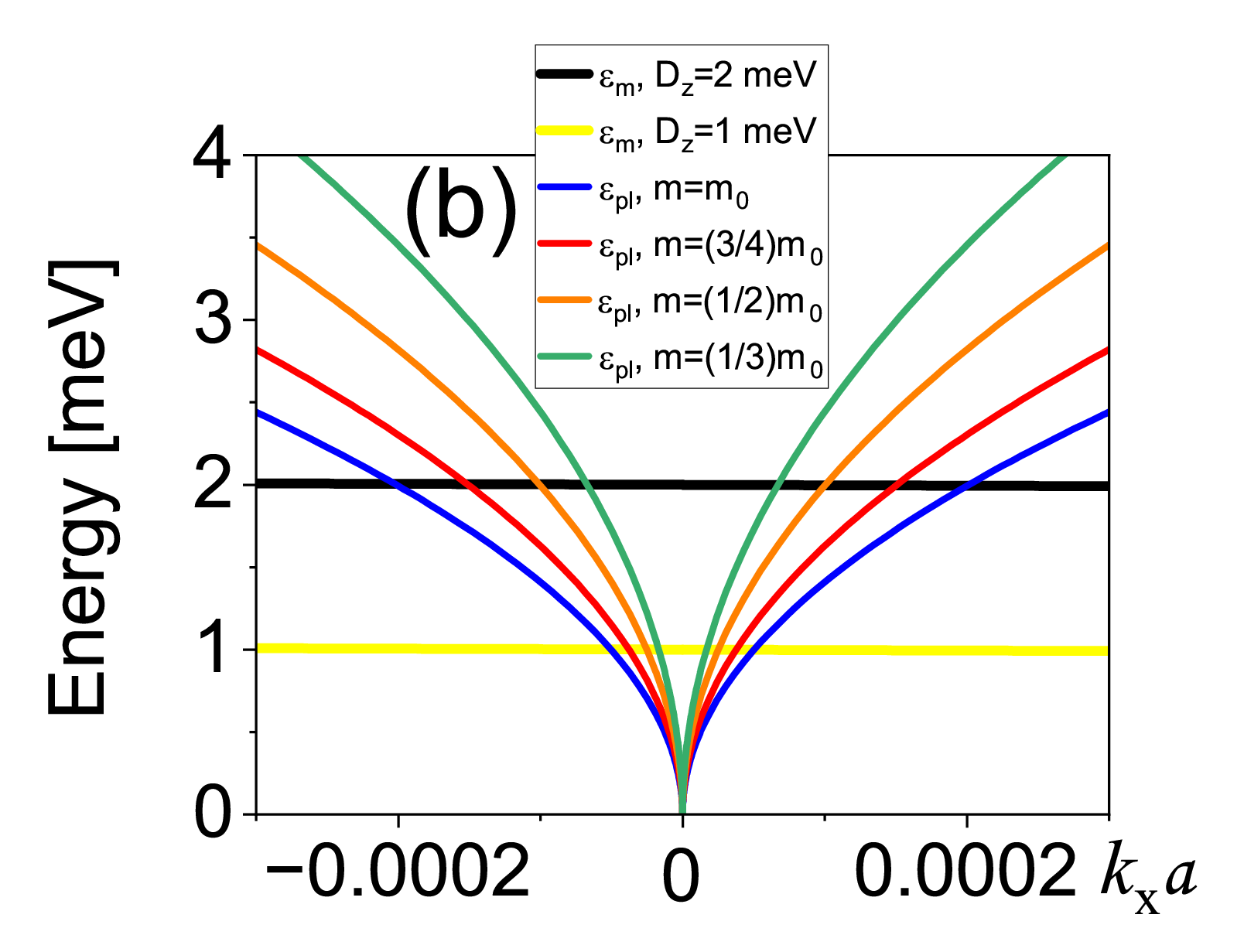}
\caption{Dispersion relations  of the uncoupled magnon and plasmon modes for the indicated values of the magnetic anisotropy constant $D_z$ and effective electron mass $m$. All other parameters are as in the main text. Behavior of the uncoupled modes  is shown in the whole Brillouin zone (a) and in the area near the Brillouin zone center (b). In the latter case the magnon curve for $D_z=1$ meV is also added to emphasize the role of magnetic anisotropy in tuning positions of the crossing points. }
\label{fig:Fig4}\end{figure}
\begin{figure*}
\includegraphics[width=1.0\textwidth]{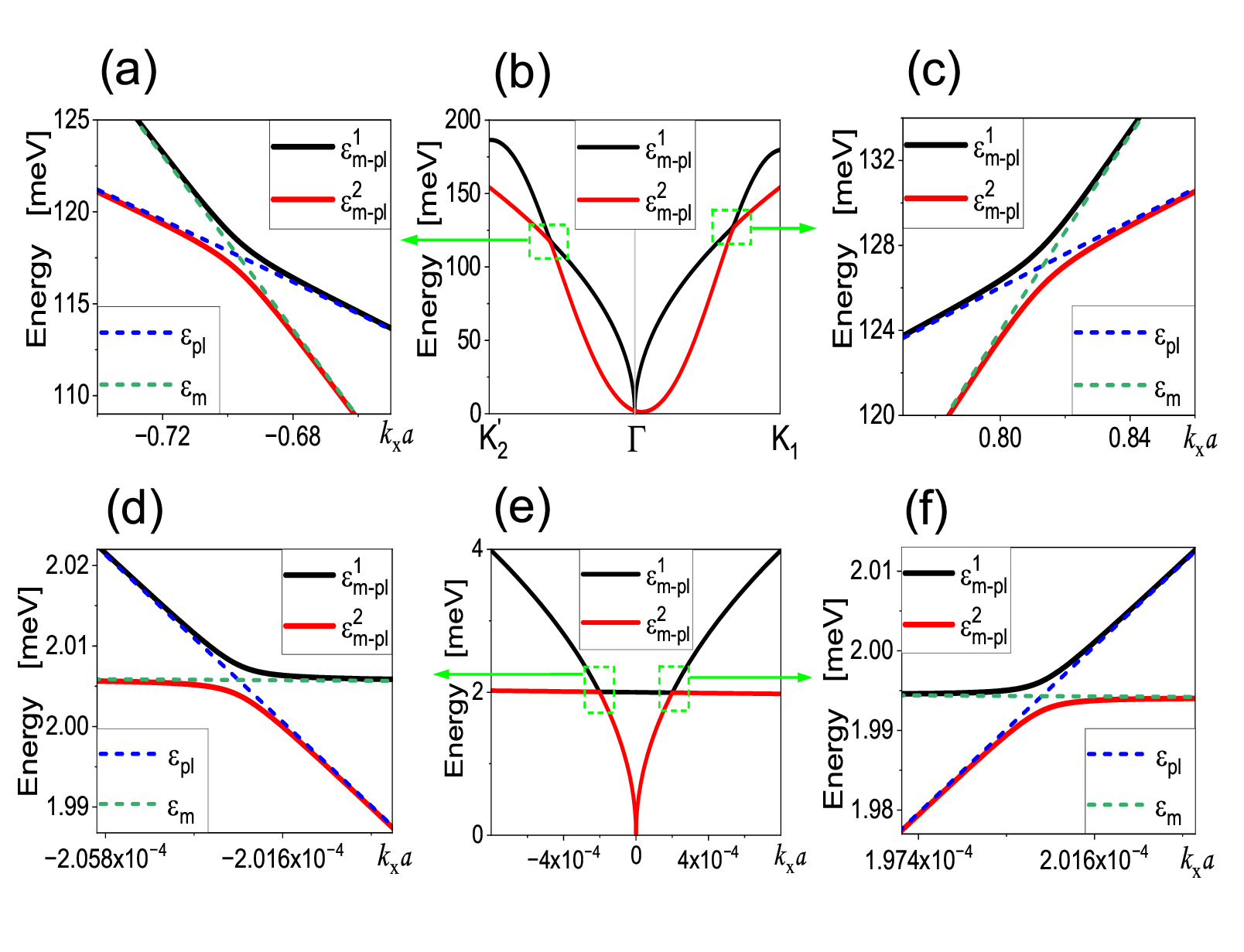}
\caption{Dispersion relations of the coupled  magnon-plasmon states (black and red solid lines),  and of the uncoupled magnon and plasmon modes (dashed green and blue lines). The upper panel (a-c) corresponds to the high energy crossings, while the lower panel (d-f) to the the crossings at low energy. The mode repulsion is clearly visible, especially for the crossing points at high energy. The parameters assumed in calculations are  $\alpha_{me}=1$, $m=m_0$,  while the other parameters are as described as in the main text.  }\label{fig:Fig5}
\end{figure*}

To show the anticrossing (mode repulsion) explicitly, we need to look in detail at the corresponding dispersion curves. To do this this,  in Fig.\ref{fig:Fig4} we present first the dispersion curves of uncoupled magnon and plasmon modes for the indicated values of the magnetic anisotropy parameter $D_z$  and effective electron mass $m$. Other parameters as in the main text. As already mentioned above, the dispersion curves of magnons and plasmons cross each other in the low energy regime, which in  Fig.\ref{fig:Fig4}(b)  takes place for $ka < 10^{-3}$. To emphasize the tuning possibility, in Fig.\ref{fig:Fig4}(a) we show the magnon dispersion relations in the whole Brillouin zone for $D_z=2$meV and indicated values of the effective mass. The area near point $\Gamma$ is enlarged in  Fig.\ref{fig:Fig4}(b), where we have also added the magnon dispersion curve for  $D_z=1$ meV -- just to emphasize that positions of the crossing points  can be tuned, as well, by tuning the the magnetic anisotropy parameter {\it via} a gate voltage. 

\begin{figure*} [t]
\includegraphics[width=0.95\textwidth]{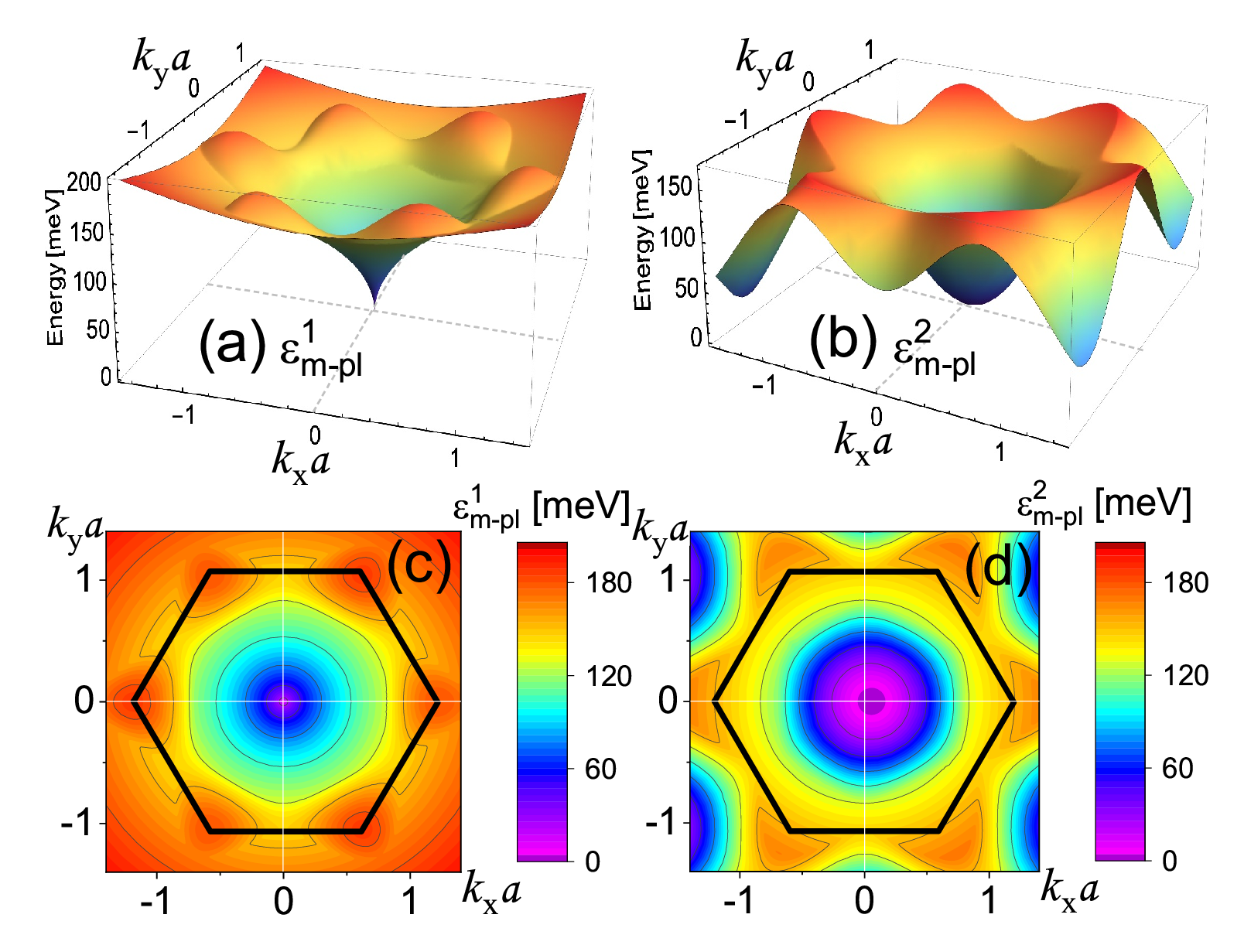}
\caption{3D plots of the upper (a) and lower  (b) energy bands of the  mixed magnon-plasmon states in the whole Brillouin zone. The corresponding 2D energy maps are plotted in (c) for the upper and (d) for the lower bands of magnon-plasmon coupled states. The parameters used are: $\alpha_{me}=1$, $m=m_0$, and other ones as in the text.}
\label{fig:Fig6}
\end{figure*}

The anticrossing behavior is shown explicitly in  Fig.\ref{fig:Fig5}, where the coupled magnon-plasmon dispersion curves are  indicated by the solid black and red lines, while the dashed lines refer to the corresponding noninteracting modes. Figure \ref{fig:Fig5} (b) shows the coupled magnon-plasmon modes in the whole Brillouin zone. As already mentioned above, there are two crossing points at high energy (one for $k>0$ and another one for $k<0$) The marked two areas near the crossing  points are zoomed in and shown in Fig.\ref{fig:Fig5}(a) for $k<0$ and  Fig.\ref{fig:Fig5}(c) for $k>0$. The mode repulsion is clearly seen in these two figures. Note, the mixed modes for $k<0$ and $k>0$ differ in energy due to DMI, and this difference is remarkable. In turn, the area near the Brillouin zone center is zoomed in and shown in Fig.\ref{fig:Fig5}(e), while the two areas near the mode crossings for $k<0$ and $k>0$ are zoomed in and shown in Fig.\ref{fig:Fig5}(d) and  Fig.\ref{fig:Fig5}(f), respectively. The anticrossing is also visible, though it is rather small due to small values of corresponding coupling parameter $c_k$. The difference in energy between the coupled modes for $k>0$ and $k<0$ is now very small and almost invisible. This appears due to a very flat magnon band in the corresponding energy window.

In Fig.\ref{fig:Fig6} we summarize our results in 3D plots of the energy of mixed magnon-plasmon modes for the upper band $\varepsilon^1_{\rm m-pl}$, Fig.\ref{fig:Fig6}(a),  and for the lower band $\varepsilon^2_{\rm m-pl}$ , Fig.\ref{fig:Fig6}(b) in the whole 3D Brillouin zone. The corresponding  2D maps are shown in  Fig.\ref{fig:Fig6}(c) and Fig.\ref{fig:Fig6}(d). Though these plots are not so transparent as those where the dispersion curves are shown explicitly, see Fig.\ref{fig:Fig5}, these  figures include all the information on the coupled magnon-plasmon states.

\section{Summary}

We have formulated a mechanism of magnon-plasmon hybridization, based on the linear magnetoelectric coupling described by a phenomenological parameter $\alpha_{me}$. The parameter $\alpha_{me}$ includes contributions from all relevant microscopic interactions. The proposed mechanism relies on the coupling of spin lattice excitations (magnons) with electric field associated with plasmons {\it via} the linear magnetoelectric effect. In a few past decades,  the magnetoelectric parameter $\alpha_{me}$ was measured in a variety of materials without inversion symmetry, and its magnitude  varies in a broad range -- from small values close to zero to maximum values close to 10, in dimensionless cgs units.
We have shown that strength of the magnon-plasmon interaction is proportional to the phenomenological parameter $\alpha_{me}$.
As the appropriate values of the parameters $\alpha_{me}$, we assumed those being somewhere in the middle of all measured parameters, and then have calculated numerically the  dispersion curves of hybridized magnon-plasmon states. On the other side, comparison of the
calculated dispersion relations with the measured ones   may be used to determine the exact value of  the magnon-plasmon coupling parameter $c_k$ and thus also of $\alpha_{me}$.

We have also considered a microscopic mechanism of magnon-plasmon coupling based on  DMI, and tuning the corresponding parameters by the electric field associated with plasmons. Unfortunately, the relevant parameters to our knowledge are  not known. The only known parameters are those describing tuning of DMI with an external electric field normal to the plane (due to gate voltage). However, these parameters on one side are irrelevant  and on the other side are too small for magnon-plasmon coupling. Therefore,  we focused calculations on the mechanism based on the phenomenological parameter $\alpha_{me}$, which effectively includes all the relevant microscopic contributions.

The proposed mechanism is different from that considered recently~\cite{Dyrdal}, where the hybridization was mediated by spin-orbit coupling associated with mobile electrons (e.g. of Rashba type).
In the present model, the spin-orbit interactions also play a key role, but they is rather associated with the lattice of localized spins.

The model calculations have been applied to the 2D VSe$_2$/NbSe$_2$ system. The proximity to NbSe$_2$ results in an out of plane magnetic anisotropy in VSe$_2$, which  in turn gives additional possibility of tuning the spin wave energy, and therefore also the hybridized magnon-plasmon modes, with a strong external electric field normal to the layer.
The magnon-plasmon  coupling can also be controlled by tuning the effective electron mass. This possibility appears quite efficient in two-dimensional systems, where the electronic band structure (and thus the effective electron mass) can be tuned externally, e.g., by strain or gate field. We believe that the magnon-plasmon hybridization in two-dimensional materials will become an important issue in the following as a link between plasmonics and magnonics.  \\

\appendix

\section {DFT calculations}

\begin{figure}
\includegraphics[width=0.9\columnwidth]{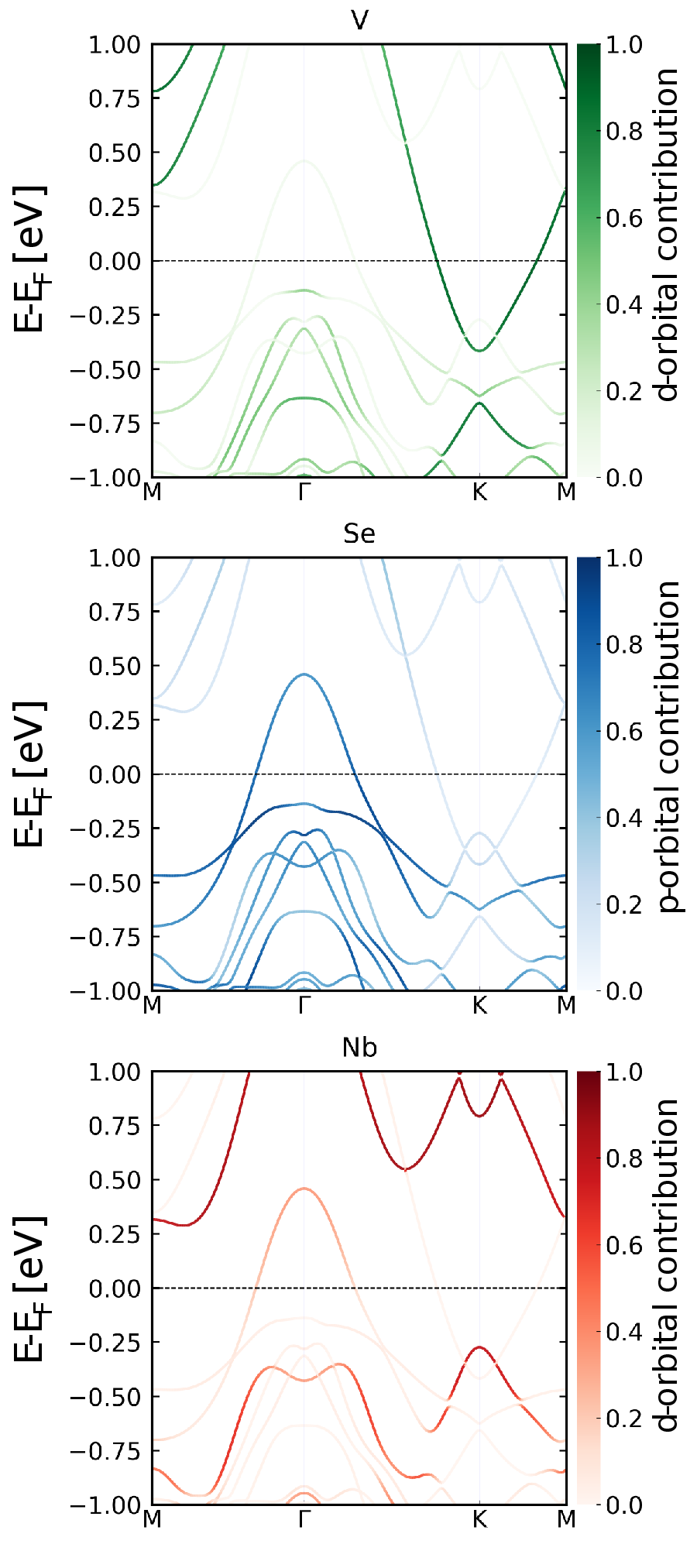}
\caption{Electronic band  structure of the VSe$_2$/NbSe$_2$ bilayer. Contributions of V (top), Se (middle) and Nb (bottom).   }
\label{fig:Fig7}
\end{figure}

We performed DFT calculations for a range of strains, and  for detailed calculations of the required parameters we selected the strain of 2\%. Arrangements of atoms in  the bilayer VSe$_2$/NbSe$_2$ is shown schematically in Fig.\ref{fig:Fig1}.The corresponding
elementary cell includes two transition metal atoms (i.e., V and Nb) and four Se atoms. The numerical calculations are based on the DFT, with the generalized gradient approximation (GGA) assumed to include exchange-correlation interactions of electrons~\cite{PhysRevLett.77.3865}. The Kohn-Sham states have been calculated using the Quantum Espresso code package, where we employed PAW pseudopotential in all calculations. The Brillouin zone was sampled using $20\times 20\times 1$ k-point grid mesh~\cite{PhysRevB.13.5188}, and the plane-wave cutoff energy was set to 60 Ry. As the structure is two-dimensional (2D), to avoid any interactions between the plane images, a 25 $\AA$ thick vacuum layer parallel to the bilayer was assumed. The total ground-state energy was calculated  with the accuracy of $10^{-9}$ eV.
Furthermore, the lattice parameters and atomic positions were optimized until the maximum force on each atom was below 0.001 eV/$\AA$. To find optimal distance between the  VSe$_2$ and NbSe$_2$ monolayers, the vdW Grimme-D3 correction~\cite{https://doi.org/10.1002/jcc.20495} was taken into account. As the unit cell includes one magnetic atom with the localized 3d-orbitals, the DFT+U was employed to consider the interaction between electrons accurately. The value of U has been set to 3 eV according to earlier studies. Similar $U$ was also assumed for Nb atoms. In all calculations, spin-orbit coupling has been included simultaneously with the Coulomb interaction.

Furthermore, to obtain the ground state of the system, we tested different spin orientations of the magnetic atoms inside the supercell, and then, from the total energy calculations, we obtained the ferromagnetic ground state of the  whole structure. To evaluate the single ion magnetic anisotropy energy (MAE), the total energy has been  computed using fully relativistic self-consistent-field DFT calculations, incorporating spin-orbit coupling (SOC) and noncollinear spin-polarization effects. The single ion MAE is defined as the difference between total energies corresponding to the magnetization orientation in-plane and out-of-plane, MAE = E[100] - E[001], and computed within the mean-field theory. Therefore, a negative (positive) value of MAE indicates an easy-plane  (easy-axis) magnetic anisotropy.
To extract the spin-spin interactions, the Hamiltonian in the basis of Wannier functions (WF) was constructed first using Wannier90.
In turn, to estimate the exchange parameters we used the minimal model spin Hamiltonian in the hexagonal lattice~\cite{LIECHTENSTEIN198765}.
All these calculations have been performed using the TB2J code package~\cite{HE2021107938}.

The corresponding band structure is shown in Fig.\ref{fig:Fig7}. From this figure follows that charge current flows predominantly via the VSe$_2$ monolayer. Therefore, the influence of the NbSe$_2$ monolayer is limited to the modification of the magnetic anisotropy and other parameters of VSe$_2$ (exchanage and DMI constants). The impact of the magnetic Nb atoms on the spin dynamics in VSe$_2$ is also ignored, as the interlayer exchange coupling is very weak, so the spin dynamics of VSe$_2$ may be considered in the first approximation as independent of that in NbSe$_2$.
The  effective parameters for the spin dynamics in VSe$_2$ are taken from the DFT calculations, and adapted to the theoretical model assumed in this paper.

\bibliography{ref.bib}

\end{document}